\newcommand{\be}{\begin{equation}}
\newcommand{\ee}{\end{equation}}

\documentclass[final
  ]
  {aipproc}

\layoutstyle{6x9}

\begin{document}

\title{Proton to pion ratio at RHIC from dynamical quark recombination}


\author{Alejandro Ayala}{
  address={Instituto de Ciencias Nucleares, Universidad
Nacional Aut\'onoma de M\'exico, Apartado Postal 70-543, M\'exico
Distrito Federal 04510, Mexico}
}

\author{Mauricio Mart\1nez}{
  address={Frankfurt Institute for Advanced Studies,
Johann Wolfgang Goethe University,
Ruth-Moufang-Str. 1
60438 Frankfurt am Main, Germany}
}

\author{Guy Pai\'c}{
  address={Instituto de Ciencias Nucleares, Universidad
Nacional Aut\'onoma de M\'exico, Apartado Postal 70-543, M\'exico
Distrito Federal 04510, Mexico} 
}

\author{G. Toledo S\'anchez}{
  address={Instituto de F\1sica, Universidad
Nacional Aut\'onoma de M\'exico, Apartado Postal 20-364, M\'exico
Distrito Federal 01000, Mexico}
}

\begin{abstract}

We propose an scenario to study, from a dynamical point of view, the
thermal recombination of quarks in the midsts of a relativistic heavy-ion
collision. We coin the term {\em dynamical quark recombination} to refer to
the process of quark-antiquark and three-quark clustering, to form mesons
and baryons, respectively, as a function of energy density. Using the {\em
string-flip model} we show that the probabilities to form such clusters
differ. We apply these ideas to the calculation of the proton and pion spectra
in a Bjorken-like scenario that incorporates the evolution of these
probabilities with proper time and compute the proton to pion ratio, comparing
to recent RHIC data at the highest energy. We show that for a standard choice
of parameters, this ratio reaches one, though the maximum is very sensitive to
the initial evolution proper time.

\end{abstract}

\maketitle


\section{Introduction}\label{I}

Recently, it has been recognized that thermal recombination of quarks plays an
important role for hadron production at intermediate $p_t$ in relativistic
heavy-ion collisions. This idea, first studied in Refs.~\cite{recomb,Fries},
explains the formation of low to intermediate $p_t$ hadrons from the bounding
of quarks in a densely populated phase space, assigning appropriate degeneracy
factors for mesons and baryons An implicit assumption is that hadronization
happens at a single temperature. However, it is known that hadronization is 
not an instantaneous process but rather that it spans a window of temperatures 
and densities. For instance lattice calculations~\cite{Karsch} show that the
phase transition from a deconfined state of quarks and gluons to a hadron gas
is, as a function of temperature, not sharp. Motivated by these shortcomings
of the original recombination scenario, here we set out to explore to what
extent the probability to recombine quarks into mesons and baryons 
depends on density and temperature and whether this probability differs for
hadrons with two and three constituents, that is to say, whether the relative
population of baryons and mesons can be attributed not only to the degeneracy
factors but rather to the dynamical properties of quark clustering in a
varying density environment.

A detailed answer to the above question stemming from first principles can only
be found by means of non-perturbative QCD. Nevertheless, in order to get a
simpler but still quantitative answer, here we address such question by
resorting to the so called string-flip model~\cite{stringflip} which has proven
to be successful in the study of quark/hadron matter as a function of
density~\cite{string1,Genaro1,Genaro2}. In this proceedings contribution, we
only outline the main features of the calculation and refer the interested
reader to Ref.~\cite{ampt} for further details. Other approaches toward a
dynamical description of recombination, in the 
context of fluctuations in heavy-ion collisions, have been recently formulated
in terms of the qMD model~\cite{0702188}. 

\section{Thermal particle spectra}\label{II}

In the recombination model, the phase space particle density is taken as the
convolution of the product of Wigner functions for each hadron's constituent
quark at a given temperature and the constituent quark wave function inside
the hadron. For instance, the meson phase space distribution is given by
\be
   F^M(x,P)=\sum_{a,b}\int_0^1dz|\Psi_{ab}^M(z)|^2w_a({\mathbf{x}},zP^+)
   \bar{w}_b({\mathbf{x}},(1-z)P^+)\, ,
   \label{wigmes}
\ee
where $P^+$ is the light-cone
momentum, $\Psi_{ab}^M(z)$ is the meson wave function and $a,\ b$ represent
the quantum  numbers (color, spin, flavor) of the constituent quark and
antiquark in the meson, respectively. An analogous equation can also be
written for baryons. When each constituent quark's Wigner function is
approximated as a Boltzmann distribution and momentum conservation is used,
the product of Wigner functions is given by a Boltzmann-like factor that
depends only on the light-cone momentum of the hadron~\cite{Fries}. For
instance, in the case of mesons 
\be
   w_a({\mathbf{x}},zP^+)\bar{w}_b({\mathbf{x}},(1-z)P^+)\sim
   e^{-zP^+/T}e^{-(1-z)P^+/T}
   =e^{-P^+/T}\, .
\ee
In this approximation, the product of parton distributions is independent of
the parton momentum fraction and the integration of the wave function over $z$
is trivially found by normalization. There can be corrections 
from a dependence of each constituent quark Wigner function on momentum
components that are not additive because energy is not conserved in this
scenario~\cite{Fries2}. An important feature to keep in mind is that in this
formalism, the QCD dynamics between quarks inside the hadron is
encoded in the wave function.  

In order to allow for a more realistic dynamical recombination scenario let us
take the above description as a guide, modifying the ingredients that account
for the QCD dynamics of parton recombination. Let us assume that the phase
space occupation can be factorized into the product of a term containing the
thermal occupation number, including the effects of a possible flow
velocity, and another term containing the system energy density $\epsilon$
driven probability ${\mathcal{P}}(\epsilon)$ of the coalescence of partons into
a given hadron. We thus write the analog of Eq.~(\ref{wigmes}) as  
\be
   F(x,P)=e^{-P\cdot v(x)/T}{\mathcal{P}}(\epsilon)\, ,
   \label{ourF}
\ee
where $v(x)$ is the flow velocity. In order to compute the probability
${\mathcal{P}}(\epsilon)$ we explicitly consider a model 
that is able to provide information about the likelihood of clustering of
constituent quarks to form hadrons from an effective quark-quark
interaction, the string-flip model, which
we proceed to describe. 

\section{String Flip Model and Hadron Recombination Probability}\label{III}

The String Flip Model is formulated incorporating a many-body quark potential
able to confine quarks within color-singlet clusters
\cite{stringflip}. At low densities, the model describes a given system of
quarks as isolated hadrons while at high densities, this system becomes a free
Fermi gas of quarks. For our purposes, we consider up and down
flavors and three colors (anticolors) quantum numbers. Our approach is very
close to that described in Refs.~\cite{string1} and~ \cite{Genaro1}, where
we refer the reader for an extensive discussion of the model details.

The many-body potential $V$ is defined as the optimal clustering of quarks into 
color-singlet objects, that is, the configuration that
minimizes the potential energy. In our approach, the interaction between
quarks is pair-wise. Therefore, the optimal clustering is achieved by finding
the optimal pairing between two given sets of quarks of different color for all
possible color charges. The minimization procedure is performed over all
possible permutations of the quarks and the interaction between quarks is
assumed to be harmonic with a spring constant $k$. Through this procedure, we
can distinguish two types of hadrons:  

i) {\it Meson-like}. In this case the pairing is imposed to be between color
and anticolors and the many-body potential of the system made up of mesons is
given by:
\be
   V_\pi = V_{B\bar{B}}+V_{G\bar{G}}+V_{R\bar{R}}\,
   \label{mespot}
\ee
where $R(\bar{R})$, $B(\bar{B})$ and $G(\bar{G})$ are the
labels for red, blue and green color (anticolor) respectively. Note that this
potential can only build pairs. 

ii) {\it Baryon-like}. In this case the pairing is imposed to be between the
different colors in all the possible combinations. In this manner, the
many-body potential is:
\be
   V_p = V_{RB}+V_{BG}+V_{RG}\, 
   \label{barpot}
\ee
which can build colorless clusters by linking 3(RBG), 6(RBGRBG),... etc.,
quarks. Since the interaction is pair-wise, the 3-quark clusters are of the
delta (triangular) shape.

The formed hadrons should interact
weakly due to the short-range nature of the hadron-hadron interaction. This is
partially accomplished by the possibility of a quark {\it flipping} from one
cluster to another. At high energy density, asymptotic freedom demands that
quarks must interact weakly. This behavior is obtained once the average
inter-quark separation is smaller than the typical confining scale.
 
We study the meson and baryon like hadrons independently. Therefore, $V=V_\pi$
or $V_p$, depending on the type of hadrons we wish to describe.
We use a variational Monte Carlo approach to describe the evolution of a
system of $N$ quarks as a function of the particle density. We consider the
quarks  moving in a three-dimensional box whose sides have length \textit{a}
and the system described by a variational wave function of the form:
\be
   \Psi_{\lambda}(\textbf{x}_1,...,\textbf{x}_N)=e^{-\lambda
   V(\textbf{x}_1,...,\textbf{x}_N)}\Phi_{FG}(\textbf{x}_1,...,\textbf{x}_N),
   \label{wavefun}
\ee
where $\lambda$ is the single variational parameter,
$V$(\textbf{x}$_1$,...,\textbf{x}$_N$) is the many-body potential either for
mesons or baryons 
and $\Phi_{FG}$(\textbf{x}$_1$,...,\textbf{x}$_N$) is the Fermi-gas wave
function given by a product of Slater determinants, one for 
each color-flavor combination of quarks. These are built up from
single-particle wave functions describing a free particle in a box
\cite{Genaro1}. 

The variational parameter has definite values for the extreme density cases.
At very low density it must correspond to the wave function solution of an
isolated hadron. For example, the non-relativistic quark model
for a hadron consisting of 2 and 3 quarks, bound by a harmonic potential,
predicts, in units where $k=m=1$ that $\lambda_\pi \to \lambda_{0\pi} =
\sqrt{1/2}$ and $\lambda_p \to \lambda_{0p} = \sqrt{1/3}$ respectively; at very
high densities the value of $\lambda$ must vanish for both cases. 

Since the simulation was performed taking $m=k=1$, to convert 
to physical units we consider each case separately.

Baryons:
To fix the the energy unit we first notice that in a 3-body system the energy
per particle, including its mass, is given by (with $m=k=1$): 
\be
\frac{E}{3}= \sqrt{3}+1.
\label{A1}
\ee
If we identify the state as the proton of mass $M_p=938$ MeV, then the
correspondence is 
\be
\sqrt{3}+1 \rightarrow 312.7\ {\mbox {MeV}}.
\label{A2}
\ee 
To fix the length unit we use the mean square radius, which for a 3-body
system is: $\sqrt{<r^2>}=(3)^{1/4}$. The experimental value for the proton
is
\be
\sqrt{<r^2>}=0.880 \pm 0.015\ {\mbox {fm}}.
\label{A3}
\ee 
Then the correspondence is: $(3)^{1/4} \rightarrow 0.88$ fm.

Mesons:
In a similar fashion we obtain for mesons (taking the pion as the
representative 2-body particle): 
Energy: $\frac{3}{2\sqrt{2}}+1 \rightarrow 70$ MeV, 
length: $2^{1/4} \rightarrow  0.764$ fm.

Our results come from simulation done with 384 particles,
192 quarks and 192 antiquarks, corresponding to having 32 $u \ (\bar{u})$ plus
32 $d\ (\bar{d})$ quarks  (antiquarks) in the three color charges
(anti-charges).  

To determine the variational parameter as a function of density we first
select the value of the particle density $\rho$ in the box, 
which, for a fixed number of particles, means changing the box size. Then we
compute the energy of the system as a function of the variational parameter
using a Monte Carlo Method. The minimum of
the energy determines the optimal variational parameter. We repeat the
procedure for a set of values of the particle densities in the region of
interest.

The information contained in the variational parameter is global, in
the sense that it only gives an approximate idea about the average size of the
inter-particle distance at a given density, which is not necessarily the same
for quarks in a single cluster. This is reflected in the behavior of the
variational parameter $\lambda_p$ for 
the case of baryons which goes above 1 for energies
close to where the sudden drop in the parameter happens. We interpret this
behavior as as a consequence of the procedure we employ to produce colorless
clusters for baryons, which, as opposed to the case to form mesons, allows the
formation of clusters with a number of quarks greater than 3. When including
these latter clusters, the information on their size is also contained in
$\lambda$. To correct for this, we compute the likelihood to find
clusters of 3 quarks $P_3$. Recall that for $3N$ quarks in the system, the
total number of clusters of 3 quarks that can be made is equal to $N$. However
this is not always the case as the density changes, given that the potential
allows the formation of clusters with a higher number of quarks. $P_3$ is 
defined as the ratio between the number of clusters of 3 quarks found at a
given density, with respect to $N$.

\begin{figure} 
  \includegraphics[height=.3\textheight]{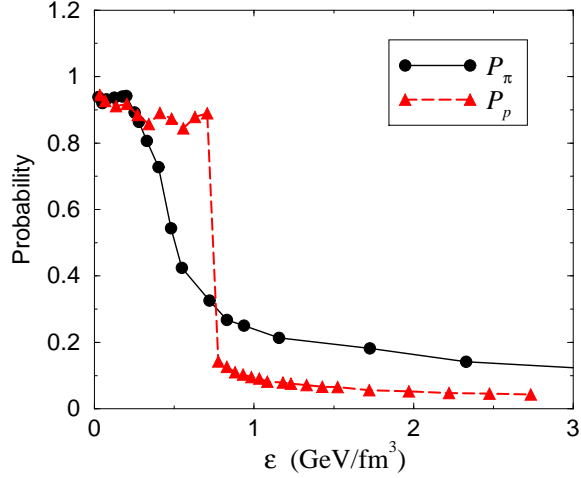}
  \caption{Probabilities to form baryons and mesons as a
  function of energy density.}
\label{P_pP_pi}
\end{figure}

Therefore, within our approach, we can define
the probability of forming a baryon as the product of the
$\lambda/\lambda_{0p}$ parameter times $P_3$, namely
\be
   {\mathcal P}_p=\lambda/\lambda_{0p} \times P_3.
   \label{probprot} 
\ee
For the case of mesons, since the procedure only takes into account the
formation of colorless quark-antiquark pairs, we simply define the probability
of forming a meson as the value of the corresponding normalized variational
parameter, namely
\be
   {\mathcal P}_\pi=\lambda/\lambda_{0\pi}.
   \label{probmes}
\ee
The probabilities ${\mathcal P}_p$ and ${\mathcal P}_\pi$ as a function of the
energy density are displayed in fig.~\ref{P_pP_pi}. Notice the qualitative
differences between these probabilities. In the case of baryons, the sudden
drop found in the behavior of the variational parameter is preserved at an
energy density around $\epsilon =0.7$ GeV/fm$^3$ whereas in the case of
mesons, this probability is smooth, indicating a difference in the production
of baryons and mesons with energy density.

%

\section{proton to pion ratio}\label{V}

In order to quantify how the different probabilities to produce sets of three
quarks (protons) as compared to sets of two quarks (pions) affect these
particle's yields as the energy density changes during hadronization, we need
to resort to a model for the space-time evolution of the collision. For the
present purposes, we will omit describing the effect of radial flow and take 
Bjorken's scenario which incorporates the fact that initially, expansion is
longitudinal, that is, along the beam direction which we take as the $\hat{z}$
axis. In this 1+1 expansion scenario, the relation between the temperature $T$
and the 1+1 proper-time $\tau$ is given by 
\be
   T=T_0\left(\frac{\tau_0}{\tau}\right)^{v_s^2},
   \label{temperaturevstau}
\ee
where $\tau=\sqrt{t^2-z^2}$. Equation~(\ref{temperaturevstau}) assumes that
the speed of sound $v_s$ changes slowly with temperature. A lattice estimate
of the speed of sound in quenched QCD~\cite{Gupta} shows that $v_s^2$
increases monotonically from about half the ideal gas limit 
for $T\sim 1.5 T_c$ and approaches this limit only for
$T>4T_c$, where $T_c$ is the critical temperature for the phase transition. No
reliable lattice results exist for the value of the speed of sound in the
hadronic phase though general arguments indicate that the equation of state
might become stiffer below $T_c$ and eventually softens as the temperature
approaches zero. For the ease of the argument, here we
take $v_s$ as a constant equal to the ideal gas limit $v_s^2=1/3$.  

We also consider that hadronization takes place on hypersurfaces $\Sigma$
characterized by a constant value of $\tau$ and therefore
\be
   d\Sigma=\tau\rho \ d\rho \ d\phi\  d\eta ,
   \label{hypersurface}
\ee
where $\eta$ is the spatial rapidity and $\rho$, $\phi$ are the polar
transverse coordinates. Thus, the transverse spectrum for a hadron species $H$
is given as the average over the hadronization interval, namely
\be
   E\frac{dN^H}{d^3P}=\frac{g}{\Delta \tau}
   \int_{\tau_0}^{\tau_f}d \tau\int_{\Sigma}d\Sigma\ \frac{P\cdot
     u(x)}{(2\pi)^3}F^H(x,P), 
   \label{distributionaveraged}
\ee
where $\Delta \tau=\tau_f-\tau_0$.

To find the relation between the energy density $\epsilon$ --that the
probability ${\mathcal{P}}$ depends upon-- and $T$, we resort to lattice
simulations. For the case of two flavors, a fair representation
of the data~\cite{Karsch} is given by the analytic expression
\be
   \epsilon /T^4 = a\left[ 1 + \tanh\left(\frac{T-T_c}{bT_c}\right)\right],
   \label{latticeenergy}
\ee
with $a=4.82$ and $b=0.132$. We take $T_c=175$ MeV.
For a purely longitudinal expansion, the flow four-velocity vector $v^\mu$ and
the normal to the freeze-out hypersurfaces of constant $\tau$, $u^\mu$,
coincide and are given by $v^\mu=u^\mu=(\cosh\eta,0,0,\sinh\eta)$,
therefore, the products $P\cdot u$ and $P\cdot v$ appearing in
Eq.~(\ref{distributionaveraged}) can be written as
\be
   P\cdot v=P\cdot u=m_t\cosh(\eta-y),
   \label{Pdotv}
\ee
where $m_t=\sqrt{m_H^2+p_t^2}$ is the transverse mass of the hadron and
$y$ is the rapidity.

\begin{figure} 
  \includegraphics[height=.3\textheight]{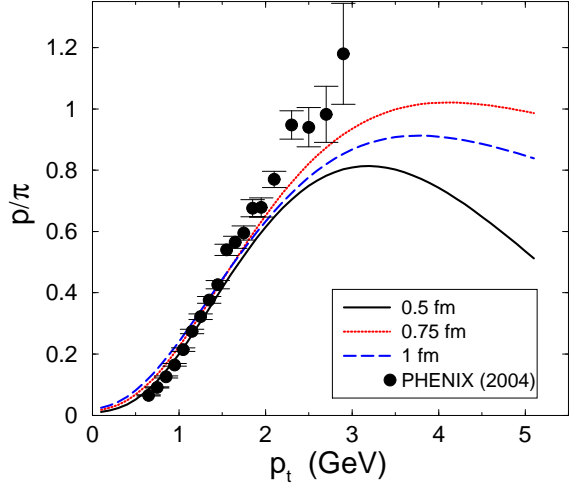}
  \caption{Proton to pion ratio as a function of transverse
  momentum for three different values of the initial evolution proper-time
  $\tau_0=0.5,\ 0.75$ and $1$ fm and the same finial freeze-out proper-time
  $\tau_f=3.5$ fm, compared to data for Au + Au collisions at
  $\sqrt{s_{NN}}=200$ GeV from PHENIX. The height of this ratio is very
  sensitive to the choice of the initial evolution time.}
\label{fig12}
\end{figure}

Considering the situation of central collisions and looking only at the case
of central rapidity, $y=0$, the final expression for the hadron's transverse
distribution is given by 
\be
   E\frac{dN^H}{d^3P}=\frac{g}{(2\pi)^3}\frac{2m_tA}{\Delta \tau }
   \int_{\tau_0}^{\tau_f}d \tau\tau K_1\left[\frac{m_t}{T(\tau )}\right]
   {\mathcal{P}}[\epsilon (\tau )].
   \label{distfin}
\ee
To obtain the the pion and proton distributions, we use the values
$\tau_0=0.75$ fm and $\tau_f=3.5$ fm and an initial temperature $T_0=200$
MeV. From Eq.~(\ref{temperaturevstau}), this corresponds to a final 
freeze-out temperature of $\sim 120$ MeV. For protons we take a degeneracy
factor $g=2$ whereas for pions $g=1$, to account for the spin degrees of
freedom. Figure~\ref{fig12} shows the proton 
to pion ratio for three different values of the initial evolution proper time
$\tau_0=0.5,\ 0.75$ and $1$ fm and the same finial freeze-out proper-time
$\tau_f=3.5$ fm, compared to data for this ratio for Au + Au collisions at
$\sqrt{s_{NN}}=200$ GeV from PHENIX~\cite{PHENIXBM}. We notice that the
maximum height reached by this ratio is sensitive to the choice of the initial
evolution time. We also notice that the $p_t$ value for which the maximum is
reached is displaced to larger values than what the experimental values
indicate. This result is to be expected since the model assumptions leading to
Eq.~(\ref{distfin}) do not include the effects of radial flow that, for a
common flow velocity, are known to be larger for protons than for pions, and
which will produce the displacement of the ratio toward lower $p_t$ values. 

\section{Summary and Conclusions}\label{concl}

In conclusion, we have used the string-flip model to introduce a dynamical
quark recombination scenario that accounts for the evolution of the
probability to form a meson or a baryon as a function of the energy
density during the collision of a heavy-ion system. We have used the model
variational parameter as a measure of the probability to form colorless
clusters of three quarks (baryons) or of quark-antiquark (mesons). We have
shown that these probabilities differ; whereas the probability to form a pion
transits smoothly from the high to the low energy density domains, the
probability to form a baryon changes abruptly at a given critical energy
density. We attribute this difference to the way the
energy is distributed during the formation of clusters: whereas for mesons the
clustering happens only for quark-antiquark pairs, for baryons the energy can
be minimized by also forming sets of three, six, etc., quarks in (colorless)
clusters. These produces competing minima in the energy that do not reach each
other smoothly. We interpret this behavior as a signal for a qualitative
difference in the probability to form mesons and a baryons during 
the collision evolution. 

We have incorporated these different probabilities to
compute the proton and pion spectra in a thermal model for a Bjorken-like
scenario. We use these spectra to compute the proton to pion ratio as a
function of transverse momentum and compare to experimental data at the
highest RHIC energies. We argue that the ratio computed from the model is able
to reach a height similar to the one shown by data, although the maximum is
displaced to larger $p_t$ values. This could be understood by recalling that
the model does not include the effects of radial flow which is known to be
stronger for protons (higher mass particles) than pions. The inclusion of
these effects is the subject of current research that will be reported
elsewhere.


\begin{theacknowledgments}
Support for this work has been received by PAPIIT-UNAM
under grant number IN116008 and CONACyT under grant number
40025-F. M. Martinez was supported by DGEP-UNAM.
\end{theacknowledgments}





\begin{thebibliography}{9}

\bibitem{recomb}
R. C. Hwa and C. B. Yang, \emph{Phys. Rev. C} \textbf{67}, 034902 (2003);
V. Greco, C. M. Ko, and P. L\'evai, \emph{Phys. Rev. Lett.} \textbf{90},
202302 (2003). 

\bibitem{Fries}
R.J. Fries, B. M\"uller, C. Nonaka and
S.A. Bass, \emph{Phys. Rev. Lett.} \textbf{90}, 202303 (2003).

\bibitem{Karsch}
F. Karsch, E. Laermann and a Peikert, \emph{Phys. Lett.} \textbf{B478},
447 (2000); F. Karsch, \emph{Lect. Notes in Phys.} \textbf{583}, 209 (2002).

\bibitem{stringflip}
C.J. Horowitz, E.J. Moniz and J.W. Negele, \emph{Phys. Rev. D} \textbf{31},
1689 (1985).

\bibitem{string1}
C. Horowitz and J. Piekarewicz, \emph{Nucl. Phys.} \textbf{A536}, 669-696
(1992). 

\bibitem{Genaro1}
G. Toledo S\'anchez and J. Piekarewicz, \emph{Phys. Rev. C} \textbf{65},
045208 (2002). 

\bibitem{Genaro2}
G. Toledo S\'anchez and J. Piekarewicz, \emph{Phys. Rev. C} \textbf{70},
035206 (2004). 

\bibitem{ampt} A. Ayala, M. Mart\1nez, G. Pai\'c and G. Toledo
S\'anchez, \emph{Dynamical quark recombination in ultrarelativistic heavy-ion
collisions and the proton to pion ratio}, arXiv:0710.3629 [hep]. 

\bibitem{0702188}
S. Haussler, S. Scherer and M. Bleicher, \emph{The effect of dynamical 
parton recombination on event-by-event observables}, hep-ph/0702188. 

\bibitem{Fries2}
R.J. Fries, B. M\"uller, C. Nonaka and
S.A. Bass, \emph{Phys. Rev. C} \textbf{68}, 044902 (2003). 

\bibitem{Gupta}
S. Gupta, Pramana {\bf 61}, 877 (2003).

\bibitem{PHENIXBM}
S.S. Adler {\it et al.} (PHENIX Collaboration), Phys. Rev. C {\bf 69},
034909 (2004).

\end{thebibliography}

\IfFileExists{\jobname.bbl}{}
 {\typeout{}
  \typeout{******************************************}
  \typeout{** Please run "bibtex \jobname" to optain}
  \typeout{** the bibliography and then re-run LaTeX}
  \typeout{** twice to fix the references!}
  \typeout{******************************************}
  \typeout{}
 }



\end{document}